# Diabetic retinopathy image classification method based on GreenBen data augmentation

Yutong Liu, Jie Gao, Haijiang Zhu*

College of Information Science and Technology, Beijing University of Chemical Technology, Beijing, 100029

**Abstract**：For the diagnosis of diabetes retinopathy (DR) images, this paper proposes a classification method based on artificial intelligence. The core lies in a new data augmentation method, GreenBen, which first extracts the green channel grayscale image from the retinal image and then performs Ben enhancement. Considering that diabetes macular edema (DME) is a complication closely related to DR, this paper constructs a joint classification framework of DR and DME based on multi task learning and attention module, and uses GreenBen to enhance its data to reduce the difference of DR images and improve the accuracy of model classification. We conducted extensive experiments on three publicly available datasets, and our method achieved the best results. For GreenBen, whether based on the ResNet50 network or the Swin Transformer network, whether for individual classification or joint DME classification, compared with other data augmentation methods, GreenBen achieved stable and significant improvements in DR classification results, with an accuracy increase of 10%.

## 1 Introduction

Diabetic retinopathy (DR) is one of the most frequent ocular complications in diabetic patients, with approximately one-third of diabetic patients developing diabetic retinopathy within five years [1,2,3]. A normal fundus image includes blood vessels, the optic disc, and the central sulcus, whereas patients with diabetic retinopathy experience different changes in the lesions as the diabetes progresses. These lesions consist of microaneurysms, soft and hard exudates, and hemorrhages. Based on these pathological features, the International Organization for the Study of Diabetic Retinopathy classifies the severity of diabetic retinopathy into five categories [4]. Diabetic Macular Edema (DME) is a diabetic ocular complication closely related to diabetic retinopathy. The pathomechanism of diabetic macular edema is complex. Hyperglycemia in diabetic patients can cause opening of the Blood Retinal Barrier (BRB) and its damage to the BRB. Once the BRB is damaged, immune cells will encounter the tissues in the retina to generate a strong immune and inflammatory response, leading to the accumulation of macromolecules in the retina and the formation of macular edema [5]. Based on the distance of the hard exudate from the center of the macula, international organizations classify the severity of diabetic macular edema into three categories [4]. The traditional diagnosis of diabetic retinopathy and diabetic macular edema relies on a comprehensive ophthalmologic examination with a complex testing process. Due to the large number of pathological features, ophthalmologists are required to have specialized knowledge and extensive experience to make an accurate diagnosis.

In recent years, Convolutional Neural Networks (CNN) have been widely used in the medical field to provide efficient and cost-effective solutions for patients. Literature [6-7] comprehensively compared CNN models such as AlexNet [8], VGGNet [9], and ResNet [10]. ResNet had the best performance, and in this paper, we will choose ResNet50 as the representative of CNN. With Transformer becoming the mainstream model for natural language processing, Vision Transformer [11] and Swin Transformer [12] models were applied in image processing. Liu [13] compared two network models, Vision Transformer and Swin Transformer, using EyePACS dataset and Swin Transformer performed better, so Swin Transformer was chosen as the representative of Transformer model. The above proposed model can categorize DR and DME separately, but ignores the intrinsic connection between them. Multi-task learning utilizes feature information from related task data, which not only improves the model performance, but also is more relevant to real life. Literature [14-16] gives the application of multi-task learning in medical images. However, for DR and DME, the lesions will be different with the progression of the disease, but the physiological structure of the retina will not change too much, and it is difficult to accomplish the fine classification task by only relying on ResNet50 and Swin Transformer. Inspired by CANet [17], CFNet [18] and the fusion of residual block and channel attention mechanism (CSAM) and non-localized block (NLB) [19], in this paper, we introduce the idiosyncrasy and dependence attention modules to enhance the detailed features of the lesion, which can localize the lesion area in the retina and measure the importance of the features according to the weights, and thus improve the feature selection performance of the model.

All the above work is dedicated to the study of the network framework, but the enhancement of the data before entering the network also has a crucial impact on the results. By applying a series of transformations to the original images, not only the size of the dataset can be increased, but also the quality of the dataset can be improved, which can improve the generalization ability of the model and reduce the generation of overfitting phenomena. In the case of fundus images, which are subject to the equipment and environment in which they are taken, noise is introduced into the fundus image and thus there are significant differences, and to eliminate such differences as much as possible and to emphasize the details of the image, some enhancement measures need to be applied to the image [20].

In view of the above problems, a new data enhancement method, GreenBen, is proposed in this paper, and to validate the effectiveness of this data enhancement method, this paper constructs a multi-task classification model for joint DME to classify DRs and incorporates an attention mechanism, i.e., superimposing idiosyncratic and dependency attention modules after the backbone network.

## 2 Materials and Supplies
### 2.1 Dataset

The paper uses three public datasets for the experiments: the Messidor [21], IDRID [22] and DeepDRID [23]. The Messidor dataset was generated from the Techno-vision project funded by the French Ministry of Defense Research in 2004, which contains a total of 1200 fundus images. The resolutions of the images are 1440×960, 2240×1488, and 2304×1536 in TIFF format. Each image was given a DR and DME severity grading label by medical experts, with four classifications for DR based on lesions such as microaneurysms and hard exudates, and three classifications for DME based on the shortest distance between the location of the hard exudate and the macular center. The IDRID dataset was obtained from the Ophthalmology Clinic, Nanded, India with a total of 516 fundus images. The images were all of resolution 4288×2848 pixels and were saved in JPG format. The training set of 413 images constitutes 80% of the total and the remaining 20% totaling 103 images constitute the test set. The entire images were graded for DR and DME by medical experts and were categorized into five categories for DR severity and three categories for DME severity according to international standards. The DeepDRID dataset is provided by the ISBI2020 Challenge and consists of the routine fundus dataset and the UWF dataset. The routine fundus dataset used in this paper consists of 2000 images from 500 patients, centered on the macula and optic disc. The resolutions of the images were 1736×1824 and 1976×1984 in JPG format, respectively. The severity of diabetic retinopathy was categorized into five categories based on international standards. All experiments were trained on data uniformly adjusted to 224*224, and three common geometric transformations for data augmentation were used: horizontal flipping, vertical flipping, and normalization.

**2.2 Experimental equipment**

The experimental equipment in this paper is Intel CoreTM i9-12900K CPU, NVIDIA A6000 GPU, Ubuntu system, modeling based on Pytorch framework. The Adam optimization algorithm is used in training, and the cross-entropy loss function is used for gradient update. The epoch is set to 1000, the batch size is 40, the initial learning rate is 3e-4, and the cosine annealing is used to attenuate the learning rate.

Taking DR as an example, the expression for the cross-entropy loss function is:

$$L_{DR} = -\frac{1}{N} \sum_{i=1}^{N} \sum_{j=1}^{M} y_i^j \log \widetilde{y_i^j} \quad （1）$$

where $L_{DR}$ is the cross-entropy function for the final classification of DR, N is the number of samples in the dataset, M is the total number of categories, M is 4 for DR and 3 for DME for the Messidor dataset, $y_i^j$ denotes the probability that sample i belongs to category j, and $\widetilde{y_i^j}$ denotes the true label of sample i.

The joint classification of DR and DME will use two cross-entropy loss functions, and the two loss functions are summed. The form is as follows:

$$L = L_{DR} + L_{DME} \quad （2）$$

Considering the effect of the idiosyncratic attention module, four cross-entropy loss functions are obtained based on the cross-entropy losses $L'_{DR}$ and $L'_{DME}$ learned from the DR and DME features, and the four loss functions are weighted and summed. The form is as follows:

$$L = L_{DR} + L_{DME} + \lambda(L'_{DR} + L'_{DME}) \quad （3）$$

where λ is the hyperparameter, set to 0.25.

## 3 Detailed Methods

### 3.1 Network architecture

The structure of the classification modeling system incorporating multi-task learning and attention mechanism used in this paper is shown in Fig. 1.

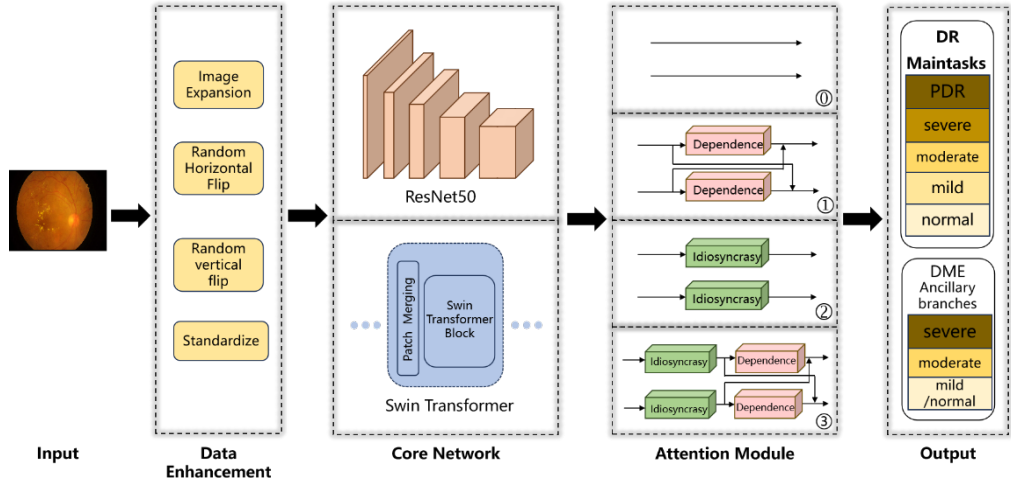

**Fig.1** System structure diagram of a classification model

The input diabetic fundus lesion images will undergo initial data preprocessing such as image deflation, random horizontal flip, random vertical flip and standardized before being passed to the backbone network. In case of individual classification, the lesion information is extracted by ResNet50 and Swin Transformer network and then directly classified to DR or DME. In case of multi-task classification, joint DME to classify DR, joint classification of DR and DME based on the features generated by the backbone network, or superimposition of dependence attention module, idiosyncrasy attention module, and combination of superimposed idiosyncrasy and dependence attention modules after the backbone network, and ultimately prediction of the classification results based on the learned idiosyncrasy and dependence features.

### 3.2 Attention module

As shown in Fig.2. The idiosyncrasy attention module uses an attention mechanism in both channel and spatial dimensions to compute the attention weights of the features to learn the feature information of the two diseases alone. The dependence attention module goes through pooling and fully connected layers to learn the feature information related to the two diseases.

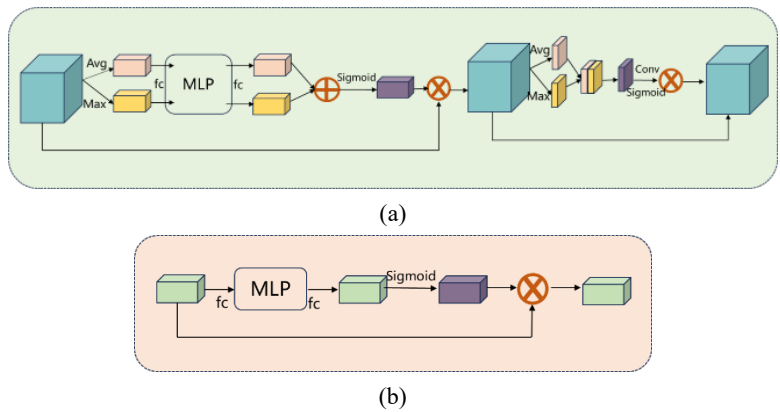

(a)

(b)

**Fig.2** (a) the idiosyncrasy attention module, (b) the dependence attention module

The idiosyncrasy attention module is constructed as follows. First, the spatial information is compressed by spatial averaging and maximum pooling operations, and then the feature information is sent to the shared MLP activated by sigmoid function, which consists of two fully connected layers. The channel attention mathematical expression is:

$$I_c = \sigma\left(MLP(P_{avg}, P_{max})\right) \quad (4)$$

The weighted features obtained from the channel attention are:

$$F_i = I_c \otimes F \quad (5)$$

Where $\otimes$ denotes the multiplication of elements in the matrix. Thus, the idiosyncrasy features of DR and DME are selected along the feature channel and suppressed for features not related to the two diseases. To further enhance the features, the weighted feature maps are then computed in spatial dimensions with the spatial attention mathematical expression:

$$I_s = \sigma\left(W(P_{avg}, P_{max})\right) \quad (6)$$

Where W and σ denote the convolution operation and sigmoid function, respectively. The feature map obtained from spatial attention is:

$$F_i' = I_c \otimes F_i \quad (7)$$

For dependent learning, the feature maps of DR and DME are used as inputs to the dependence attention module after average pooling and fully connected operations, and the MLP and sigmoid functions are used for learning the attention weights. The output expression of the dependence attention module is:

$$G_{DR}' = F_{DR}' \oplus F_{DME}' \otimes \sigma\left(MLP\left(F_c(P_{avg}(F_{DME}'))\right)\right) \quad (8)$$

Where $F_{DR}'$ and $F_{DME}'$ denote the idiosyncrasy outputs of DR and DME, respectively, and $F_c$ and $P_{avg}$ denote the fully connected and average pooling operations, respectively. As a result, the network can learn the relationship between two diseases, DR and DME, and improve the effect of joint DME on DR classification.

### 3.3 Data enhancement

For fundus images, the green channel is often considered the most useful because the wavelength of green light penetrates the retina better, making the details of structures such as blood vessels more visible[24].Ben Graham proposed a new method of image enhancement in the Diabetic Retinopathy Competition [25]. First the fundus image is subjected to a Gaussian blurring operation, which reduces the noise in the fundus image and smooths the image details to simulate the background of the fundus image. Then the Gaussian blurred fundus image is subtracted from the original fundus image to obtain a differentiated image to enhance the edge and texture features in the fundus image. Contrast Limited Adaptive Histogram Equalization (CLAHE) [26] uses linear interpolation to optimize the transition problem between blocks and reduce the problem of AHE noise amplification. At the same time CLAHE by limiting the increase of image contrast, the image will become more harmonious and the subtle structures in the image will become clearer.

Therefore, in this paper, we try to extract the green channel grayscale image from the RGB image to enhance the contrast of the image by reducing the redundant features. Although the use of green channel grayscale images can improve the performance of image analysis, maintaining multi-channel information may provide more contextual information, which can help the model to better understand the image content and perform image classification. Considering that Ben image enhancement enhances the effective features of the image by reducing the background expression, try to put the gray scale image after extracting the green channel and then go through Ben image enhancement, and find that better classification results can be obtained, naming this data enhancement method as GreenBen enhancement.

The image details of the original image after CLAHE, Ben Image Enhancement, and GreenBen Image Enhancement processes respectively are shown in Figure 3.

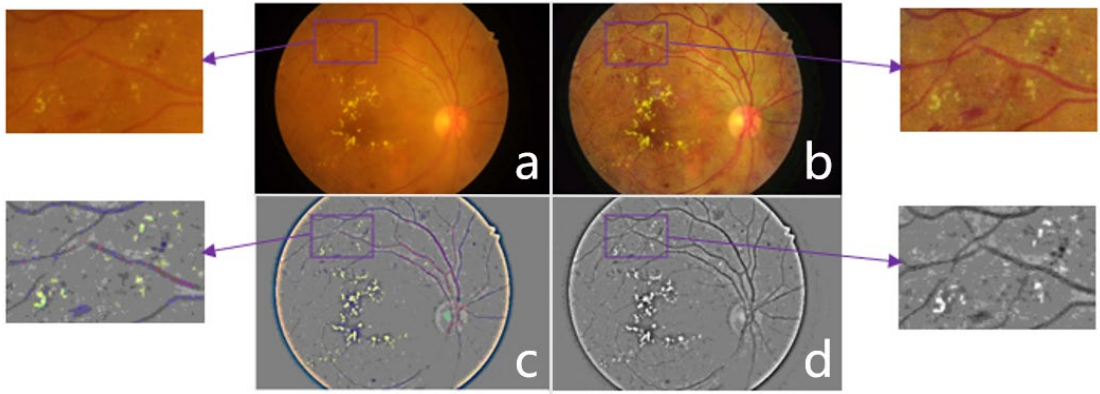

**Fig.3** (a) original image, (b) CLAHE, (c) Ben enhancement, (d) GreenBen enhancement images details

### 3.4 Evaluation indicators

In this paper, the accuracy, area under the ROC curve (AUC), precision, recall and F1-score were used to evaluate the experimental results. To measure the joint grading effect of DR and DME, Joint Accuracy metrics (Joint Acc) were used. The formulas for these metrics are as follows, respectively:

$$A_{CC} = \frac{TP+TN}{TP+TN+FN+FP} \quad (9)$$

$$P_{re} = \frac{TP}{TP+FP} \quad (10)$$

$$R_{ec} = \frac{TP}{TP+FN} \quad (11)$$

$$F_1 = \frac{2TP}{2TP+FN+FP} \quad (12)$$

Where TP denotes the number of samples correctly judged as positive by the model, FP denotes the number of samples in which the model incorrectly judged negative samples as positive, TN denotes the number of negative samples correctly judged by the model, and FN denotes the number of samples in which the model failed to identify correctly and incorrectly judged positive samples as negative samples. Meanwhile, to measure the effect of joint grading of DR and DME, the Joint Accuracy proposed by IDRID Challenge was used, and the calculation rule is that if the model predicts the labeling of DR and DME for both pairs of fundus images, the result is counted as 1, otherwise it is counted as 0, and the final number of samples with 1 is compared to the total number of images.

### 3.5 Network architecture
#### 3.5.1 Classification results under multi-task learning and attentional mechanisms

ResNet50 and Swin Transformer classify DR and DME individually (RDR, RDME, SDR, SDME);

classify DR jointly with DME (ResNet50 and Swin); superimpose after the backbone network the dependence attention module (ResNet1 and Swin1), the idiosyncrasy attention module (ResNet2 and Swin2), and combined superimposed idiosyncrasy and dependence attention modules (ResNet3 and Swin3). The experimental results on the Messidor dataset are shown in Table 1, and on the IDRID dataset are shown in Table 2, with bolded numbers indicating the best results.

**Table 1** The performance of different models on the Messidor dataset

| Model | | DR | | | | | DME | | | | |
|---|---|---|---|---|---|---|---|---|---|---|---|
| | Joint Acc | Acc | AUC | Pre | Rec | F1 | Acc | AUC | Pre | Rec | F1 |
| RDR | — | 0.719 | 0.856 | 0.639 | 0.628 | 0.622 | — | — | — | — | — |
| RDME | — | — | — | — | — | — | 0.843 | **0.961** | 0.653 | 0.803 | 0.695 |
| ResNet50 | 0.838 | 0.913 | 0.955 | 0.905 | 0.883 | 0.893 | 0.914 | 0.905 | 0.804 | 0.716 | 0.739 |
| ResNet1 | **0.849** | **0.917** | 0.955 | **0.928** | 0.867 | 0.895 | **0.920** | 0.927 | **0.805** | 0.726 | 0.750 |
| ResNet2 | 0.842 | 0.911 | **0.958** | 0.893 | 0.893 | 0.892 | 0.919 | 0.907 | **0.805** | 0.725 | 0.743 |
| ResNet3 | 0.848 | **0.917** | 0.956 | 0.897 | **0.902** | **0.899** | **0.920** | 0.923 | 0.797 | **0.750** | **0.759** |
| SDR | — | 0.521 | 0.672 | 0.352 | 0.384 | 0.354 | — | — | — | — | — |
| SDME | — | — | — | — | — | — | 0.830 | 0.728 | 0.379 | 0.398 | 0.382 |
| Swin | 0.679 | 0.753 | 0.777 | 0.808 | 0.535 | 0.625 | 0.867 | 0.793 | 0.653 | 0.527 | 0.542 |
| Swin1 | 0.678 | 0.752 | 0.786 | 0.711 | 0.557 | 0.622 | 0.861 | 0.799 | 0.619 | 0.516 | 0.533 |
| Swin2 | 0.634 | 0.693 | 0.715 | 0.774 | 0.323 | 0.424 | 0.836 | 0.707 | 0.403 | 0.405 | 0.391 |
| Swin3 | 0.588 | 0.618 | 0.605 | 0.362 | 0.172 | 0.208 | 0.813 | 0.580 | 0.304 | 0.338 | 0.307 |

**Table 2** Performances of different models on the IDRID dataset

| Model | | DR | | | | | DME | | | | |
|---|---|---|---|---|---|---|---|---|---|---|---|
| | Joint Acc | Acc | AUC | Pre | Rec | F1 | Acc | AUC | Pre | Rec | F1 |
| RDR | —— | 0.553 | 0.840 | 0.376 | 0.428 | 0.397 | —— | —— | —— | —— | —— |
| RDME | —— | —— | —— | —— | —— | —— | 0.796 | **0.911** | 0.661 | 0.641 | 0.645 |
| ResNet50 | 0.553 | 0.602 | 0.776 | 0.459 | 0.451 | **0.451** | 0.786 | 0.856 | 0.654 | 0.632 | 0.637 |
| ResNet1 | **0.573** | 0.602 | **0.802** | 0.453 | 0.433 | 0.426 | **0.835** | 0.787 | **0.892** | **0.696** | **0.728** |
| ResNet2 | 0.553 | 0.583 | 0.785 | 0.457 | 0.444 | 0.442 | 0.806 | 0.879 | 0.684 | 0.647 | 0.654 |
| ResNet3 | 0.553 | **0.641** | **0.802** | 0.587 | 0.454 | 0.434 | 0.806 | 0.881 | 0.668 | 0.648 | 0.652 |
| SDR | —— | 0.534 | 0.777 | 0.400 | 0.374 | 0.367 | —— | —— | —— | —— | —— |
| SDME | —— | —— | —— | —— | —— | —— | 0.777 | 0.864 | 0.638 | 0.600 | 0.597 |
| Swin | 0.524 | 0.544 | 0.697 | 0.449 | 0.379 | 0.371 | 0.757 | 0.823 | 0.625 | 0.586 | 0.583 |
| Swin1 | 0.524 | 0.544 | 0.697 | 0.449 | 0.379 | 0.371 | 0.757 | 0.823 | 0.625 | 0.586 | 0.583 |
| Swin2 | 0.563 | 0.602 | 0.771 | 0.426 | 0.419 | 0.407 | 0.796 | 0.839 | 0.545 | 0.589 | 0.561 |
| Swin3 | 0.524 | 0.544 | 0.714 | 0.514 | 0.378 | 0.359 | 0.748 | 0.828 | 0.498 | 0.552 | 0.524 |

Messidor and IDRID datasets, ResNet50 improves classification accuracy (DR Acc) by 19% and 5% over RDR, and Swin improves classification accuracy (DR Acc) by 23% and 1% over SDR, which shows that multi-task learning can effectively improve the classification effect. Messidor and IDRID datasets are classified on ResNet50 The classification effect is optimal after overlaying the idiosyncrasy and dependence attention module, but this attention module is not applicable on Swin Transformer, probably because Swin Transformer still has a large receptive field although it divides the local window, while DR foci are more biased towards local features.

**3.5.2 Classification results with data enhancement**

To validate the impact of the GreenBen data enhancement method proposed in this paper when classifying DR individually, the DeepDRID dataset is used in this paper for validation. To validate its impact when classifying DR jointly, this paper uses IDRID dataset for validation. The experimental results of using several data enhancement methods such as GreenBen and CLAHE on DeepDRID and IDRID datasets on this paper's model are given in Table 3 and Table 4, respectively. The GreenClahe method of extracting the green channel before CLAHE enhancement is also explored in this paper.

Table 3 The performance of different data augmentation methods in the DeepDRID dataset

| Method | ResNet2 | | | | | Swin2 | | | | |
|---|---|---|---|---|---|---|---|---|---|---|
| | Acc | AUC | Pre | Rec | F1 | Acc | AUC | Pre | Rec | F1 |
| None | 0.753 | 0.890 | 0.666 | 0.570 | 0.581 | 0.613 | 0.799 | 0.370 | 0.406 | 0.374 |
| CLAHE | 0.720 | 0.903 | 0.632 | 0.606 | 0.606 | 0.608 | 0.810 | 0.469 | 0.412 | 0.397 |
| Ben | 0.778 | **0.921** | 0.660 | **0.643** | 0.629 | 0.643 | **0.829** | 0.476 | 0.466 | 0.463 |
| Green Channel | 0.765 | 0.911 | 0.627 | 0.614 | 0.582 | 0.565 | 0.806 | 0.362 | 0.368 | 0.351 |
| GreenClahe | 0.755 | 0.918 | 0.618 | 0.591 | 0.595 | 0.618 | **0.829** | 0.451 | 0.437 | 0.428 |
| GreenBen | **0.783** | **0.921** | **0.688** | 0.620 | **0.632** | **0.653** | 0.821 | **0.479** | **0.471** | **0.473** |

DeepDRID dataset after CLAHE, Ben enhancement, extracting green channel and GreenClahe method on ResNet2 model has -3%, 2.5%, 1%, 0.2% improvement in classification accuracy (Acc) compared to no data enhancement, but GreenBen enhancement method proposed in this paper has 3% improvement.Swin2 model There are -0.5%, 3%, -5%, 0.5%, and 4% enhancements in the classification accuracy of each enhancement method on the Swin2 model. The effectiveness of the GreenBen enhancement method in classifying DR alone under ResNet50 and Swin Transformer is fully demonstrated.

Table 4 The performance of different data augmentation methods in the IDRID dataset

| Method | ResNet2 | | | | | ResNet3 | | | | |
|---|---|---|---|---|---|---|---|---|---|---|
| | Acc | AUC | Pre | Rec | F1 | Acc | AUC | Pre | Rec | F1 |
| None | 0.583 | 0.785 | 0.457 | 0.444 | 0.442 | 0.641 | 0.802 | **0.587** | 0.454 | 0.434 |
| CLAHE | 0.621 | 0.784 | 0.430 | 0.437 | 0.422 | 0.631 | 0.835 | 0.478 | 0.466 | 0.451 |
| Ben | 0.612 | **0.838** | 0.463 | 0.450 | 0.448 | 0.592 | 0.778 | 0.497 | 0.461 | 0.460 |
| Green Channel | 0.650 | 0.812 | **0.521** | 0.486 | **0.469** | 0.631 | 0.826 | 0.522 | **0.471** | 0.461 |
| GreenClahe | 0.641 | 0.822 | 0.514 | 0.486 | 0.482 | **0.650** | **0.851** | 0.507 | 0.459 | 0.445 |
| GreenBen | **0.689** | 0.793 | 0.493 | **0.488** | 0.468 | 0.641 | 0.804 | 0.486 | **0.471** | **0.466** |

IDRID dataset after CLAHE, Ben enhancement, extracting green channel and GreenClahe method on ResNet2 model has 4%, 3%, 7%, 6% improvement in classification accuracy (Acc) compared to no data enhancement, but GreenBen enhancement method proposed in this paper has 10% improvement. on ResNet3 model GreenBen enhancement is only slightly worse than GreenClahe, and all other data enhancement methods produce a decrease in accuracy (Acc). The stability of the GreenBen method is significantly better than that of GreenClahe, which again proves the effectiveness of the GreenBen enhancement method.

**4 Potential Pitfalls and Trouble Shooting**

Our method GreenBen is optimal for classification results under both classification models on the DeepDRID dataset. However, on the IDRID dataset, GreenBen still outperforms several other common enhancements of classification results on the classification model with the addition of the idiosyncratic attention module. However, after adding the dependent attention module, GreenBen no longer has the leading edge, and even the effect of some other data enhancement methods is reduced. The reason may be that the classification model after adding the dependent attention module needs to classify according to some texture features in the fundus image, but the GreenBen enhancement method happens to weaken this part of subtle feature information as noise information, resulting in a limited enhancement of the final classification index.

**5 Conclusion**

In order to reduce the influence of noise in fundus images on the classification results of diabetic retinopathy, this paper proposes the GreenBen data enhancement method, which is experimented on a classification model incorporating multi-task learning and attention mechanism using DeepDRID and IDRID datasets, and its results show that both under the condition of separate classification and under the condition of joint classification with joint DME The GreenBen method is effective, and the classification accuracy can be improved by up to 10%. Moreover, the GreenBen enhancement method is applicable to both ResNet50 and Swin Transformer network structures.

**References**
[1] Teo Z L, Tham Y C, Yu M C Y, et al. Global prevalence of diabetic retinopathy and projection of burden through 2045: systematic review and meta-analysis[J]. Ophthalmology, 2021, 128(11): 1580-1591.
[2] Zhang J S. Progress in treatment of diabetes retinopathy[J]. Journal of Hubei Minzu University (Medical Edition), 2021, 38(01): 88-90.
[3] Antonetti D A, Silva P S, Stitt A W. Current understanding of the molecular and cellular pathology of diabetic retinopathy[J]. Nature Reviews Endocrinology,2021,17(4):195-206.
[4] Flaxel C J, Adelman R A, Bailey S T, et al. Diabetic Retinopathy Preferred Practice Pattern [J]. Ophthalmology, 2020, 127(1): P66-66P145.
[5] Shin E S, Sorenson C M, Sheibani N. Diabetes and retinal vascular dysfunction[J]. Ophthalmic Vis Res, 2014, 9(3): 362-373.
[6] Li X G, Pang T T, Xiong B, et al. Convolutional neural networks based transfer learning for diabetic retinopathy fundus image classification[J]. In: Proceedings of the 10th International Congress on Image and Signal Processing, BioMedical Engineering and Informatics (CISP-BMEI). Shanghai, China: IEEE,

2017:1−11.

[7] Wan S H, Liang Y, Zhang Y. Deep convolutional neural networks for diabetic retinopathy detection by image classification[J]. Computers & Eletrical Engineering, 2018, 72: 274-282.

[8] Alex K, Ilya S and Geoffrey E. H. ImageNet classification with deep convolutional neural networks[C]// Proceedings of the 25th International Conference on Neural Information Processing Systems,2012:1097-1105.

[9] Simonyan K and Andrew Z. Very deep convolutional networks for large-scale image recognition[J]. Computer science, 2014(3):11-18.

[10] K He, X. Zhang, S Ren and J Sun. Deep residual learning for image recognition[C]// Proceeding of the IEEE conference on computer vision and pattern recognition. 2016: 770-778.

[11] Vaswani, Ashish, Noam M. et al. Attention is All you Need[J] .Neural Information Processing Systems (2017).

[12] Alexey D, Lucas B, Alexander K, et al. An image is worth 16x16 words: transformers for image recognition at scale[J]. https://arxiv.org/abs/2010.11929 (2020): 1-22.

[13] LIU Y Y. Research on detection of diabetic retinopathy based on deep learning[D].Shenyang University of Technology, 2022.

[14] XING G, MIAO Z, ZHENG Y, et al. A multi-task model for reliable classification of thyroid nodules in ultrasound images [J]. Biomedical engineering letters,2024,14(2):187-197.

[15] TAN Z, MADZIN H, NORAFIDA B, et al. DeepPulmoTB: A benchmark dataset for multi-task learning of tuberculosis lesions in lung computerized tomography (CT)[J].Heliyon, 2024,10(4):e25490.

[16] SHAO W, SHI H, LIU J, et al. Multi-instance Multi-task Learning for Joint Clinical Outcome and Genomic Profile Predictions from the Histopathological Images [J]. IEEE transactions on medical imaging,2024.

[17] LI X M,HU X W,YU L Q,et al.CANet:cross-disease attention network for joint diabetic retinopathy and diabetic macular edema grading[J].IEEE Transactions on Medical Imaging,2020,39(5):1483-1493.

[18] LI Z F, WU L P, LI J J. Diagnosis of diabetic retinopathy based on context fusion network[J]. Journal of Jiangsu Ocean University (Natural Science Edition), 2023, 32(01): 51-62.

[19] Madarapu S, Ari S, Mahapatra K. A deep integrative approach for diabetic retinopathy classification with synergistic channel-spatial and self-attention mechanism[J].Expert Systems With Applications,2024,249(PA):123523.

[20] FAN J W, ZHANG R R, LU M. Applications of Deep Learning Techniques for Diabetic Retinal Diagnosis[J]. Acta Automatica Sinica,2021,47(05): 985-1004.

[21] Decenciere E, Zhang X,Cazuguel G, et al. Feedback on a publicly distributed image database: the messidor database[DB/OL]. Image Analysis & Stereology, 33(03): 231-234,2014.

[22] P. Porwal, S. Pachade, R. Kamble, et al. Indian diabetic retinopathy image dataset(idrid): A dataset for diabetic retinopathy screening research[DB/OL]. Data, 3(03): 25,2018.

[23] Liu R,Wang X, Wu Q, et al. Deepdrid: Diabetic retinopathy——grading and image quality estimation challenge[DB/OL]. Patterns, 100512, 2022.

[24] Datta N S, Dutta H S, Majumder K. Brightness-preserving fuzzy contrast enhancement scheme for the detection and classification of diabetic retinopathy disease[J]. Journal of Medical Imaging, 2016, 3(1):

014502.

[25] Ben Graham. Kaggle diabetic retinopathy detection competition report [EB/OL]. 2019.

[26] Mira H, Kahlil M, Roslidar, et al. Impact of CLAHE-based image enhancement for diabetic retinopathy classification through deep learning[J].Procedia Computer Science,2023,21657-66.